\documentclass[runningheads]{llncs}

\usepackage{natbib}
\usepackage{graphicx}
\usepackage{amsmath}
\usepackage{amssymb}

\begin{document}

\title{Cardiac Functional Analysis with Cine MRI via Deep Learning Reconstruction}

\author{Eric Z. Chen\inst{1} \and Xiao Chen\inst{1} \and Jingyuan Lyu\inst{2} \and Qi Liu\inst{2} \and Zhongqi Zhang\inst{3} \and Yu Ding\inst{2} \and Shuheng Zhang\inst{3} \and Terrence Chen\inst{1} \and Jian Xu\inst{2} \and Shanhui Sun\inst{1}}

\institute{United Imaging Intelligence, Cambridge, MA, USA \and UIH America, Inc., Houston, TX, USA \and United Imaging Healthcare, Shanghai, China}

\authorrunning{E. Chen et al.}
\titlerunning{ }% Part of RIGHT running header

\maketitle

\begin{abstract}
Retrospectively gated cine (retro-cine) MRI is the clinical standard for cardiac functional analysis. Deep learning (DL) based methods have been proposed for the reconstruction of highly undersampled MRI data and show superior image quality and magnitude faster reconstruction time than CS-based methods. Nevertheless, it remains unclear whether DL reconstruction is suitable for cardiac function analysis. To address this question, in this study we evaluate and compare the cardiac functional values (EDV, ESV and EF for LV and RV, respectively) obtained from highly accelerated MRI acquisition using DL based reconstruction algorithm (DL-cine) with values from CS-cine and conventional retro-cine.
To the best of our knowledge, this is the first work to evaluate the cine MRI with deep learning reconstruction for cardiac function analysis and compare it with other conventional methods. The cardiac functional values obtained from cine MRI with deep learning reconstruction are consistent with values from clinical standard retro-cine MRI.

%\keywords{Real-time cine  \and MRI image reconstruction \and Deep learning.}
\end{abstract}

%\section{Synopsis}
%% up-to 100 word
%Retrospectively gated cine (retro-cine) MRI is the clinical standard for cardiac functional analysis.
%To the best of our knowledge, this is the first work to evaluate the cine MRI with deep learning reconstruction for cardiac function analysis and compare it with other conventional methods. The cardiac functional values obtained from cine MRI with deep learning reconstruction are consistent with values from clinical standard retro-cine MRI.
% 100 words

%\section{Summary of Main Findings}
%  250 Characters (~35 words)
%A 250 character summary of the abstract’s main findings is requested for generation of digital poster previews. It is expected that this content will be overlap with the 100-word synopsis. As with the synopsis, summary must include text only, without equations or images and be without references or citations to items described in the full abstract.

%This is the first work to evaluate the cine MRI with deep learning reconstruction for cardiac function analysis and compare it with other conventional methods. The cardiac functional values obtained from cine MRI with deep learning reconstruction are consistent with values from clinical standard retro-cine MRI.
 
% 237 Characters

\section{Introduction}
% INTRODUCTION: “Why was this study/research performed? What unsolved problem are you addressing?”

Cardiac functional analysis is crucial for the diagnosis and treatment of cardiovascular diseases (CVD) \citep{karamitsos2009role}. It often involves measuring the end-diastolic volume (EDV), end-systolic volume (ESV) and ejection fraction (EF) for the left and right ventricles from a series of cardiac images. Conventional segmented, retrospectively gated cine (retro-cine) MRI, the gold standard for cardiac MR functional analysis, provides dynamic information of the heart motion that is critical for function evaluation \citep{moon2002breath}. Multiple slices (e.g., 8-10 short-axis slices) are typically needed to achieve whole heart coverage and therefore multiple breath-holdings are required. Despite its wide clinical use, retro-cine may suffer from poor image quality resulting from the inconsistent breath-holding position of the patient, prolonged exam time due to its segmented acquisition, and limitation of use in patients challenged by the attempt to hold their breaths.

Compressed sensing (CS) is an emerging fast MR imaging technique utilizing the theory of compressed sensing with a specifically designed acquisition and reconstruction strategy \citep{jaspan2015compressed}. The potential of CS-cine in cardiac functional analysis with the highly undersampled acquisition (8X-15X acceleration) has been demonstrated \citep{feng2013highly}. However, its clinical use is hampered by long reconstruction time due to the nature of iterative algorithms. This challenge could be further amplified in the reconstruction of multiple slices (e.g., one-hour reconstruction time \citep{feng2013highly}).

Deep learning (DL) based methods have been proposed for the reconstruction of highly undersampled MRI data and show superior image quality and magnitude faster reconstruction time than CS-based methods \citep{chen2020real}. Nevertheless, it remains unclear whether DL reconstruction is suitable for cardiac function analysis. To address the above question, in this study we evaluate and compare the cardiac functional values (EDV, ESV and EF for LV and RV, respectively) obtained from highly accelerated MRI acquisition using DL based reconstruction algorithm (DL-cine) with values from CS-cine and conventional retro-cine. To the best of our knowledge, this is the first work to evaluate the cine MRI with DL reconstruction for cardiac function analysis.

\section{Methods}

\textbf{Data collection}: We collected total 216 cine MRI data from seven volunteers and one patient with cardiomegaly using a bSSFP sequence on a clinical 3T scanner (uMR 790 United Imaging Healthcare, Shanghai, China) with the approval of local IRB. 
For each subject, we acquired three datasets with different protocols and reconstruction methods, namely, retro-cine, CS-cine,  DL-cine. Each dataset includes nine short-axis slices.  

The imaging parameters for retro-cine were: imaging matrix: 224 x 199 x 9(slices), TR/TE = 3.13/1.48 ms, FOV = 360 x 320 mm$^2$, slice thickness = 8 mm, flip angle = 43°$\sim$50°, bandwidth = 1200 Hz/pixel, temporal resolution was 34ms, reconstructed cardiac phase=25. The k-space was prospectively undersampled with net reduction factor of 1.8. The imaging parameters for CS-cine and DL-cine were: imaging matrix: 224 x 199 x 9(slices), TR/TE = 2.89/1.34 ms, bandwidth = 1200 Hz/pixel, 15 phase encoding lines were collected for each cardiac phase, with a temporal resolution of 43.4 ms, FOV = 360 x 320 mm$^2$, slice thickness = 8 mm, flip angle = 50°.The k-space was prospectively undersampled using a lookup table using variable Latin Hypercube Sampling \citep{Lyu2019toward} with a net reduction factor of 15.

\textbf{Image reconstruction}: 
For retro-cine, parallel imaging was used for reconstruction.
For CS-cine, temporal sparsity regularization was used in the CS reconstruction framework.
For DL-cine, we adopted the convolutional recurrent neural network model, Res-CRNN \citep{chen2020real}, which takes into account the dynamic information and reconstruct cardiac cine images in an iterative fashion. The model was pre-trained using retro-cine data \citep{chen2020real} and directly applied to the cine data in this study. %For RT-cine (CS) and retro-cine, the CS-based reconstruction and GRAPPA were used, respectively.

\textbf{Cardiac functional analysis}: To delineate the LV epi- and endo-cardium and RV epi-cardium, a UNet-based segmentation model \citep{ronneberger2015u} was first used to provide preliminary LV myocardium, LV blood pool and RV blood pool segmentation. The model was trained on retro-cine data with ground truth segmentation provided by ACDC \citep{bernard2018deep}, and applied to all collected data in this study. The images with the overlayed segmentation were then randomized and provided blindly to an experienced researcher to undergo manual screening and adjustment. LV and RV volumes were then calculated from the multi-slice segmentation using Simpson's rule. EDV, ESV were obtained as the maximum and minimum points on the volume curve, respectively. EF was calculated as $EF = (EDV-ESV)/ESV $.

% 301 words

\section{Results}
Figure 1 shows the examples of the reconstructed images from DL-cine, CS-cine and retro-cine MRI at ES and ED. All methods show good image quality. Figure 2 shows the average values of EF, EDV and ESV for LV and RV across different methods. No statistically significant difference was found between DL-cine and retro-cine for EF, EDV and ESV (all p$>$0.05 by Wilcoxon signed-rank test). CS-cine also shows similar results with retro-cine except for lower average values of LV EF than retro-cine (p $<$ 0.05). The Bland-Altman plots (Figure 3 and 4) indicate that DL-cine has a good agreement with retro-cine and it is better than CS-cine since a few values are out of 95$\%$ limits of agreement in the latter. 

\section{Discussion and Conclusion}
%The average EF and EDV for LV is lower in RT-cine than retro-cine MRI \cite{zhang2014real,kowallick2014real, feng2013highly}
% 71 words

This is the first work that compares cine MRI using DL reconstruction with retro-cine for cardiac functional analysis. In this study, for EF, EDV and ESV, no statistically significant difference was found between DL-cine and retro-cine MRI, which is the clinical standard for cardiac functional analysis. 
Compared to retro-cine, DL-cine has the  advantage of faster data acquisition. Furthermore, DL-cine has faster reconstruction speed as well as superior image quality than CS-cine \citep{chen2020real}. 
This suggests that cine MRI with DL reconstruction is suitable for clinical usage on cardiac functional analysis with fast data acquisition and image reconstruction, which can be especially beneficial for patients who have difficulties holding their breath during the scan.

 \begin{figure}[htbp]
\centering
\includegraphics[width=\linewidth]{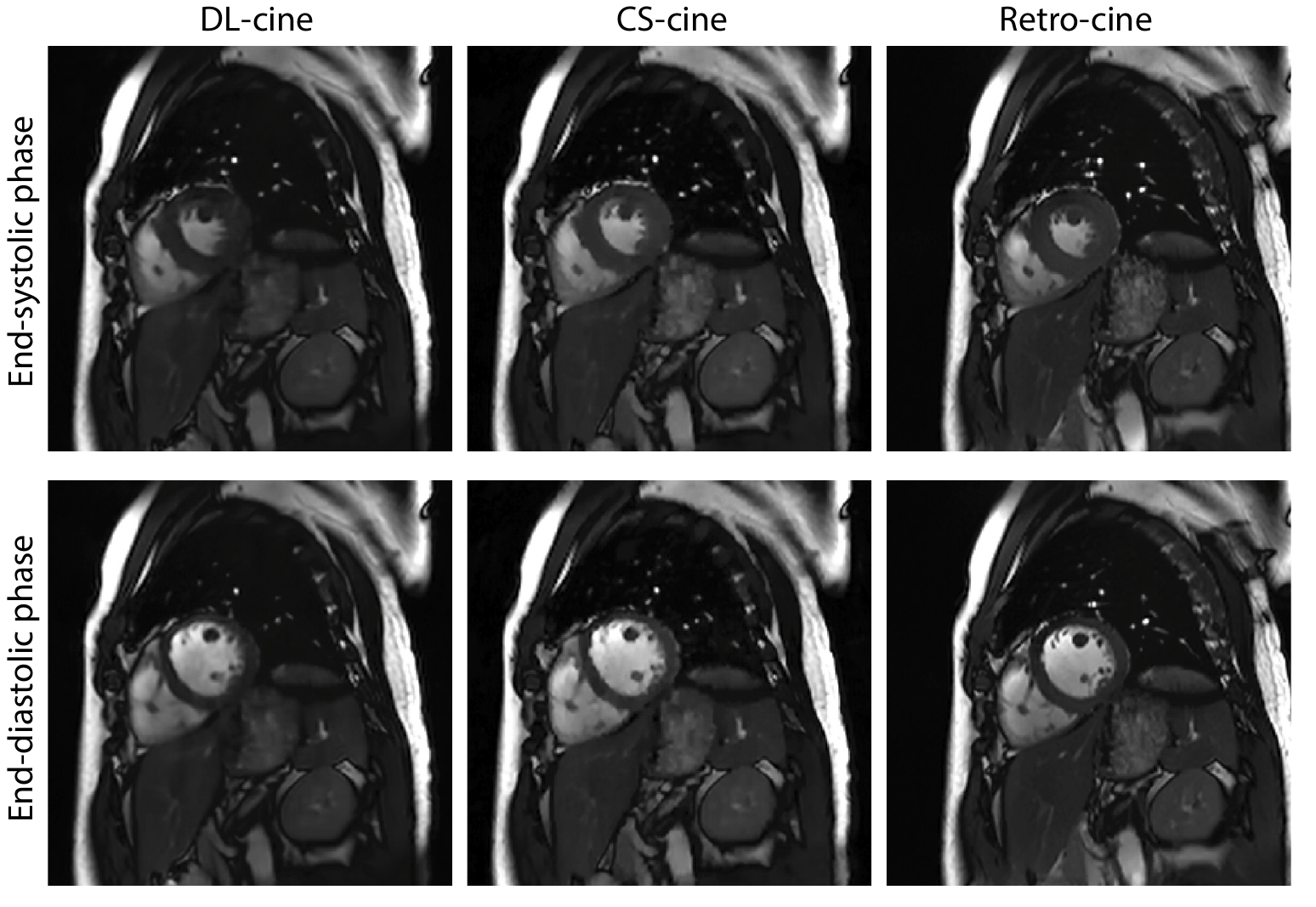}
\caption{Examples of reconstructed images from DL-cine, CS-cine and retro-cine MRI. Data were acquired from the same subject but three different scans.}
\label{fig:recon_image}
\end{figure}

\begin{figure}[htbp]
\centering
\includegraphics[width=\linewidth]{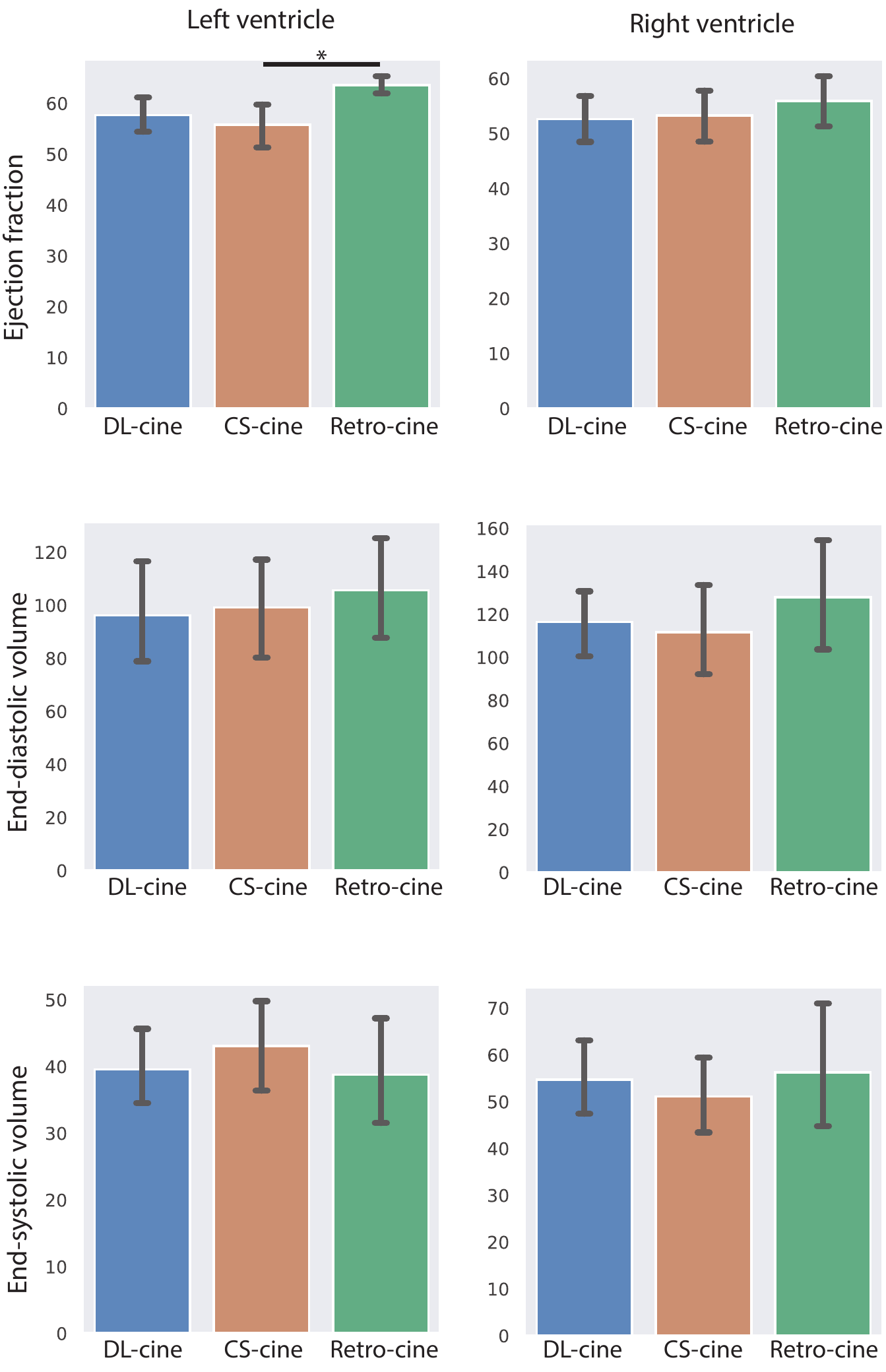}
\caption{Cardiac functional analysis based on DL-cine, CS-cine and retro-cine. Difference with statistical significance (p$<$0.05) is indicated by the star.
}
\label{fig:cardiac_values}
\end{figure}

\begin{figure}[htbp]
\centering
\includegraphics[width=\linewidth]{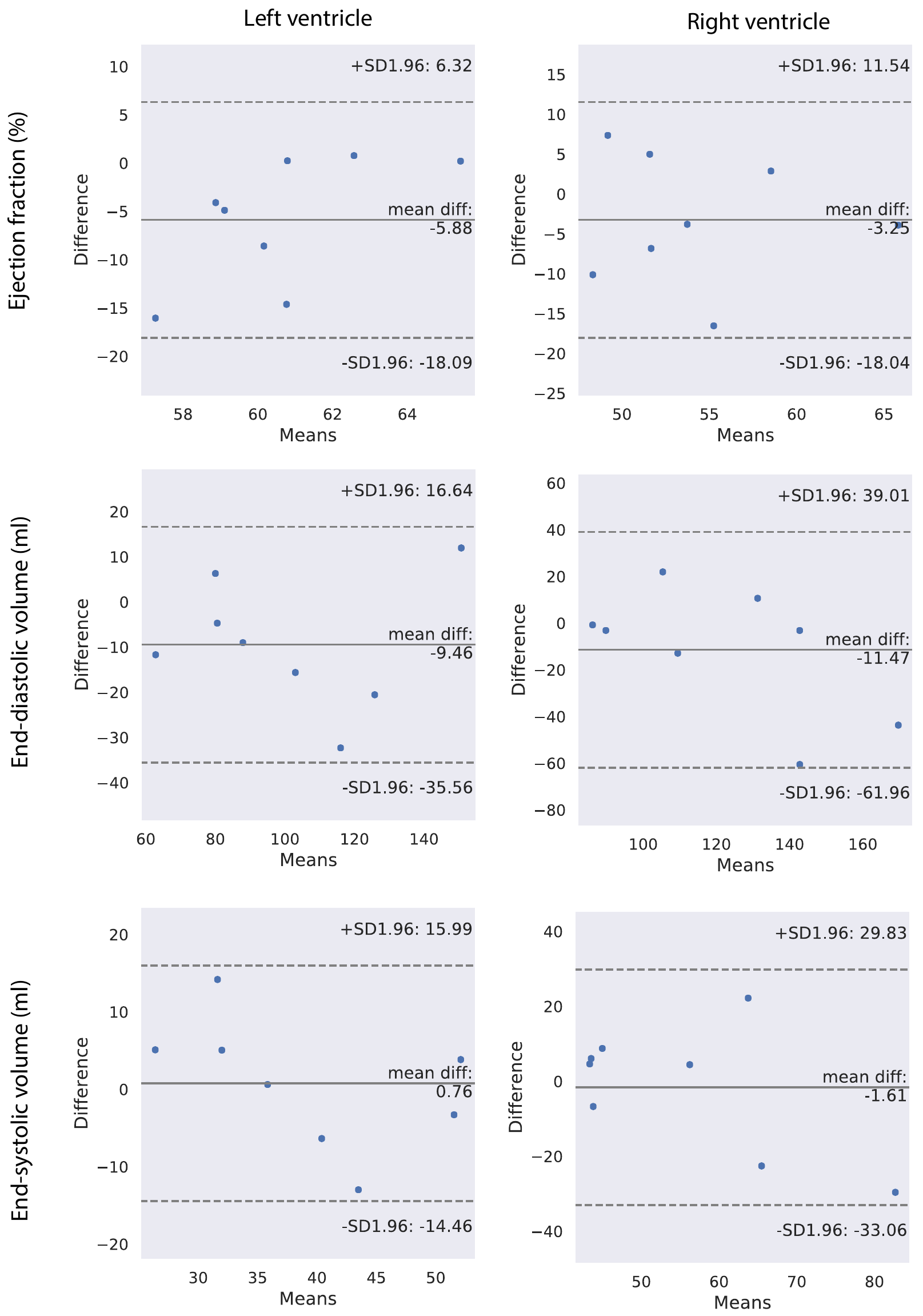}
\caption{Bland-Altman plots comparing the cardiac functional values between DL-cine and retro-cine. The solid line is the mean difference and the dash lines indicate lower and upper 95$\%$ limits of agreement. All values are within the 95$\%$ limits of agreement. 
}
\label{fig:altman_dl_vs_retro}
\end{figure}

\begin{figure}[htbp]
\centering
\includegraphics[width=\linewidth]{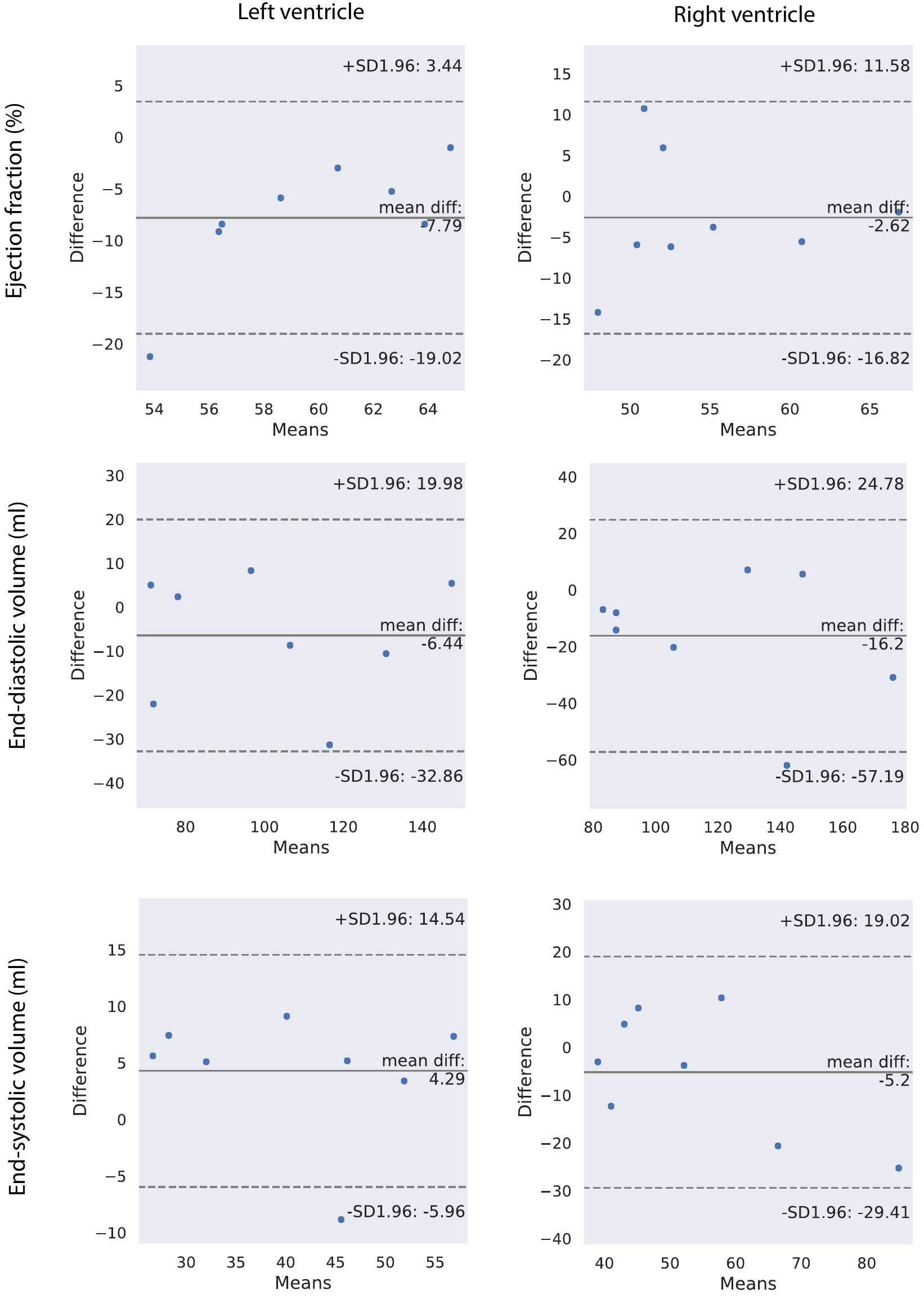}
\caption{Bland-Altman plots comparing the cardiac functional values between CS-cine and retro-cine. The solid line is the mean difference and the dash lines indicate lower and upper 95$\%$ limits of agreement. Most of the values are within the 95$\%$ limits of agreement.
%Difference with marginally statistical significance (p=0.04) is indiciated by the stars.
}
\label{fig:altman_cs_vs_retro}
\end{figure}

%\begin{figure}[htbp]
%\centering
%\includegraphics[width=\linewidth]{Fig/altman_dl_vs_cs_crop.pdf}
%\caption{Bland-Altman plots comparing the cardiac functional values between RT-cine (DL) and RT-cine (CS). %Difference with marginally statistical significance (p=0.04) is indiciated by the stars.
%}
%\label{fig:altman_dl_vs_cs}
%\end{figure}

% 93 words

%\bibliographystyle{ieeetr}
%\bibliographystyle{plain}
%\bibliographystyle{unsrt}
%\bibliographystyle{splncs04}
%\bibliographystyle{unsrtnat}
\bibliographystyle{abbrvnat}

\bibliography{references}
\end{document}